\documentclass[preprintnumbers,showkeys,showpacs,byrevtex]{revtex4}
\usepackage{amsmath,amsfonts,amssymb,amscd,amsxtra,amsthm}
\usepackage{graphicx}
\usepackage{multirow}
\usepackage{bm}
\usepackage{epstopdf}
\begin{document}
\preprint{KIAS-P12040}
\title{Electrical  conductivity of quark matter at finite $T$ under external magnetic field}
%--------------------------------------------------
\author{Seung-il Nam}
\email[E-mail: ]{sinam@kias.re.kr}
\affiliation{School of Physics, Korea Institute for Advanced Study (KIAS), Seoul 130-722, Republic of Korea}
%--------------------------------------------------
\date{\today}
\begin{abstract}
We investigate the electrical  conductivity $(\sigma)$ of quark matter via the Kubo formula at finite temperature and zero quark density $(T\ne0,\mu=0)$ in the presence of an external strong magnetic field. For this purpose, we employ the dilute instanton-liquid model, taking into account its temperature modification with the trivial-holonomy caloron distribution. By doing that, the momentum and temperature dependences for the effective quark mass and model renormalization scale are carefully evaluated. From the numerical results, it turns out that $\sigma\approx(0.02\sim0.15)\,\mathrm{fm}^{-1}$ for $T=(0\sim400)$ MeV with the relaxation time $\tau=(0.3\sim0.9)$ fm. In addition, we  also parameterize the electrical conductivity as $\sigma/T\approx (0.46,0.77,1.08,1.39)\,C_\mathrm{EM}$ for $\tau=(0.3,0.5,0.7,0.9)$ fm, respectively. These results are well compatible with other theoretical estimations, including those from the lattice QCD simulations. It also turns out that the external magnetic field plays only a minor role for $\sigma$ even for the very strong one $B_0\sim m^2_\pi\times10$ and becomes relatively effective for $T\lesssim200$ MeV. Moreover, we compute the soft photon emission rate from the quark-gluon plasma, using the electrical  conductivity calculated. 
\end{abstract}
\pacs{13.40.-f,12.39.-x,11.10.Wx,12.38.Mh}
\keywords{Electrical  conductivity, quark matter, finite temperature, instanton, caloron.}
\maketitle
%--------------------------------------------------
\section{Introduction}
%--------------------------------------------------
QCD at extreme conditions has been one of the most celebrated subjects in high-energy physics for decades. The observation of the strong magnetic field $B_0\sim m^2_\pi$ in the non-central heavy-ion collision at Relativistic Heavy Ion Collider (RHIC) of BNL~\cite{Voloshin:2008jx,Voloshin:2004vk,AA:2009txa,AA:2009uh} triggered abundant research works, such as the chiral magnetic effect (CME)~\cite{Kharzeev:2004ey,Kharzeev:2007jp,Buividovich:2009wi,Fukushima:2008xe}, chiral magnetic wave (CMW)~\cite{Kharzeev:2010gd,Burnier:2011bf}, QCD phase structure in the presence of magnetic fields, related to the magnetic catalysis~\cite{Menezes:2008qt,Nam:2011vn,Bali:2011qj}, electromagnetic properties of QCD matter~\cite{Buividovich:2010tn,Ding:2010ga,Kerbikov:2012vp,Osipov:2007je,Hiller:2008eh,Sadooghi:2012wi}, and so on. In addition to the magnetic field, the transport coefficients for the hot and/or dense matter is of great importance in general as well, since they characterize the physical properties of the matter. In general, those coefficient can be studied by the Kubo formula~\cite{BOOKEM,Gupta:2003zh,Karsch:2007jc}. Among them, in the present work, we want to investigate the electrical  conductivity $(\sigma)$, which corresponds to the vector-current correlation (VCC) and its spectral functions~\cite{Ding:2010ga}, for quark matter at finite temperature and zero quark density $(T\ne0,\mu=0)$. This condition resembles the present heavy-ion collision experimental conditions, such as RHIC and Large Hadron Collider (LHC) of CERN, in the presence of the external magnetic field. We note that the conductivity effects much on the magnetic field in terms of the relaxation time~\cite{Buividovich:2010tn,Kerbikov:2012vp}. The spectral functions are deeply related to the thermal dilepton production, and the production cross section contains much information on the hadron properties, produced in quark-gluon plasma (QGP)~\cite{Harada:2006hu}.

The electrical conductivity was investigated in the hot phase of the QCD plasma and extracted from a quenched SU(3) lattice QCD (LQCD) simulation of the Euclidean time VCC for $1.5\le T/T_c \le3$, resulting in $\sigma/T\approx7C_\mathrm{EM}$, where $C_\mathrm{EM}=\sum_fe^2_f$ and $e_f$ denotes the electric charge for the quark with a flavor $f$, in Ref.~\cite{Gupta:2003zh}. Similarly, the electrical conductivity was studied with quenched SU(3) LQCD with the maximum entropy method (MEM) for VCC in the deconfined phase $T\sim1.45\,T_c$ in Ref.~\cite{Ding:2010ga}, in which the dilepton production rate was studied as well. In that work, the electrical conductivity turned out to be $1/3\lesssim\sigma/(C_\mathrm{EM}T)\lesssim1$. Note that the estimated value for $\sigma$ is about one order larger in Ref.~\cite{Gupta:2003zh} in comparison to those from Ref.~\cite{Ding:2010ga}. In Ref.~\cite{Buividovich:2010tn}, VCC in quenched SU(2) LQCD simulation was investigated with a chirally invariant lattice Dirac operator with a constant external magnetic field. It was observed that $\sigma_\perp$ and $\sigma_\parallel$, which are perpendicular and parallel to the external magnetic field, respectively, are almost the same and insensitive to the external magnetic field for $T\gtrsim T_c$, whereas there appear considerable differences between them at $T\sim0$ with respect to the magnetic field. The electrical conductivity was also estimated as $\sigma_\perp\approx\sigma_\parallel\approx\sigma=(0.076\pm0.010)\,\mathrm{fm}^{-1}$ for $T\approx350$ MeV with $T=1.12\,T_c$. Beside the LQCD simulations, using the Green-function method, the electrical conductivity was explored at finite $T$ and $\mu$: $T\approx100$ MeV and $\mu\approx400$ MeV~\cite{Kerbikov:2012vp}. At specific choices for the model parameters, the electrical conductivity was resulted in $\sigma\approx0.04\,\mathrm{fm}^{-1}$, which is similar to those from Refs.~\cite{Buividovich:2010tn,Ding:2010ga}. Interestingly, they found that the electrical conductivity is very insensitive to the external magnetic field.

Considering the rather unsettled situation for the electrical conductivity for hot and dense quark matter as mentioned above, in the present work, we employ the dilute instanton-liquid model to study it~\cite{Shuryak:1981ff,Diakonov:1983hh}. The model is characterized by the two phenomenological (instanton) parameters: average inter-(anti)instanton distance $\bar{R}\approx1$ fm and (anti)instanton size $\bar{\rho}\approx1/3$ fm. In the instanton ensemble, representing the nonperturbative QCD vacuum, the quarks are delocalized by flipping their helicities. As a result, the quarks acquire their momentum-dependent effective mass, corresponding to the finite value for the chiral condensate and manifesting the spontaneous breakdown of chiral symmetry (SBCS). Moreover, the momentum-dependent quark mass plays the role of a UV regulator by construction. Since we are interested in the case at finite $T$, the caloron with trivial holonomy is used to modify the instanton parameters~\cite{Diakonov:1988my,Nam:2009nn}. By doing that, the instanton size $\bar{\rho}$ becomes a smoothly decreasing function of $T$, signaling weakening nonperturbative effects of QCD. In addition, we utilize the fermionic Matsubara formula for inclusion of $T$ in the relevant physical quantities in hand. In order to take into account the external magnetic field, which is produced in the peripheral heavy-ion collisions, the Schwinger method is employed~\cite{Schwinger:1951nm,Nieves:2006xp,Nam:2008ff}. 

Using the Kubo formula for the electrical conductivity, we compute $\sigma$ as a function of $T$ and the relaxation time $\tau$ for the SU(2) light-flavor sector in the chiral limit, i.e. $m_u\sim m_d\sim0$. From the numerical results, it turns out that $\sigma\approx(0.02\sim0.15)\,\mathrm{fm}^{-1}$ for $T=(0\sim400)$ MeV with the relaxation time $\tau=(0.3\sim0.9)$ fm. In addition, we  also parameterize the electrical conductivity as $\sigma/T\approx (0.46,0.77,1.08,1.39)\,C_\mathrm{EM}$ for $\tau=(0.3,0.5,0.7,0.9)$ fm, respectively. These results are well compatible with other theoretical estimations, including those from the lattice QCD simulations. It also turns out that the external magnetic field plays only a minor role for $\sigma$ even for the very strong magnetic field $B_0\approx m^2_\pi\times10$ and becomes relatively effective for $T\lesssim200$ MeV. Moreover, we also compute the soft photon emission rate from the quark-gluon plasma, using the electrical  conductivity calculated. 

The present work is organized as follows: In Section II, we briefly introduce the present theoretical framework. The numerical results and related discussions will be given in Section III. The last Section is devoted to summary, conclusion, and future perspectives.

%--------------------------------------------------
\section{Theoretical framework}
%--------------------------------------------------
Now, we make a brief explanation for the theoretical framework for computing the electrical  conductivity. We note that all the calculations are performed in flavor SU(2) in the chiral limit $(m_u,m_d)\to0$. First, the electrical  conductivity can be defined in Euclidean space from the Kubo formula as follows~\cite{Kerbikov:2012vp}:
%EQUATION>>>
\begin{equation}
\label{eq:EC1}
\sigma_{\mu\nu}(p)=-\sum_f\frac{e^2_f}{w_p}
\int\frac{d^4k}{(2\pi)^4}\mathrm{Tr}_{c,\gamma}
\left[S(k)\gamma_\mu S(k+p)\gamma_\nu\right]_A.
\end{equation}
%EQUAITON<<<
Here, $e_f$ stands for the electrical  charge for a light-flavor ($f$) quark, i.e. $(e_u,e_d)=(2e/3,-e/3)$, where $e$ denotes the unit charge $\sqrt{4\pi\alpha_\mathrm{EM}}$ with the fine structure constant $\alpha_\mathrm{EM}\approx1/137$. $w_p$ indicates the Matsubara frequency for the momentum $p$ for $\sigma$, being proportional to $2\pi T$. $\mathrm{Tr}_{c,\gamma}$ is assigned as the trace over the color and Lorentz indices. The subscript $A$ at $[\cdots]_A$ means that we have introduced the externally induced electromagnetic (EM) vector field $A_\mu$. Note that, in Eq.~(\ref{eq:EC1}), we have taken into account only the connected-diagram contribution, since the disconnected-diagram one $\propto(e_u+e_d)^2$ contributes to the electrical  conductivity negligibly~\cite{Ding:2010ga}. The longitudinal and transverse components for the electrical conductivity are defined with Eq.~(\ref{eq:EC1}) as
%EQUATION>>>
\begin{equation}
\label{eq:SIGMA}
\sigma_\perp\equiv\lim_{w_p\to0}\lim_{\bm{p}\to0}[\sigma_{11,22}(p)],\,\,\,\,
\sigma_\parallel\equiv\lim_{w_p\to0}\lim_{\bm{p}\to0}[\sigma_{33}(p)].
\end{equation}
%EQUAITON<<<

In order to evaluate Eq.~(\ref{eq:EC1}), we make use of the dilute instanton-liquid model~\cite{Shuryak:1981ff,Diakonov:1983hh}. This model is based on the idea that the quarks interact with the nontrivial QCD-vacuum configuration, i.e. instanton ensemble, by flipping their chiralities. As a result, the quark acquires the dynamically generated effective mass, and spontaneously breakdown of chiral symmetry (SBCS) is explained. The model is also characterized by the two phenomenological parameters for the light-flavor sectors: average inter-(anti)instanton distance $\bar{R}$ and average (anti)instanton size $\bar{\rho}$. For vacuum, these values are phenomenologically chosen to be $(\bar{R},\bar{\rho})\approx(1,1/3)$ fm~\cite{Shuryak:1981ff,Diakonov:1983hh}. In Euclidean space, the effective chiral action (EChA) of the model reads:
%EQUATION>>>
\begin{equation}
\label{eq:ECA}
\mathcal{S}_\mathrm{eff}=-\mathrm{Sp}_{c,f,\gamma}\ln
\left[i\rlap{\,/}{D}-iM(D^2) \right],
\end{equation}
%EQUAITON<<<
where $\mathrm{Sp}_{c,f,\gamma}$ stands for the functional trace over the color, flavor, and Lorentz indices, while $M(D^2)$ for the momentum-dependent effective quark mass from the quark zero-mode solution in the instanton vacuum~\cite{Diakonov:1983hh}. Note that $D$ indicates the U(1) covariant derivative $iD_\mu=i\partial_\mu+e_fA_\mu$. In this way, we induce the external electromagnetic (EM) field to EChA: the Schwinger method~\cite{Schwinger:1951nm,Nieves:2006xp,Nam:2008ff}. From EChA, one can derive the light-quark propagator under the external EM field in the momentum space as~\cite{Nam:2009jb}
%EQUATION>>>
\begin{equation}
\label{eq:PRO}
S(K)=\frac{\rlap{\,/}{K}+iM^2_K}
{K^2+M^2_K}\approx
\frac{\rlap{\,/}{K}+i[M_k+\frac{1}{2}\tilde{M}_k(\sigma\cdot F)]}{k^2+M^2_k}.
\end{equation}
%EQUAITON<<<
Here, $K_\mu\equiv k_\mu+e_fA_\mu$ and $\sigma\cdot F\equiv \sigma_{\mu\nu}F^{\mu\nu}$. $F^{\mu\nu}$ stands for the photon field strength tensor. The effective quark mass and its derivative $\tilde{M}_k$ in Eq.~(\ref{eq:PRO}) are defined by~\cite{Nam:2009jb}:
%EQUATION>>>
\begin{equation}
\label{eq:MASS1}
M_k=M_0\left[\frac{2}{2+\bar{\rho}^2k^2} \right]^2,\,\,\,\,
\tilde{M}_k=-\frac{8M_0\bar{\rho}^2}{(2+\bar{\rho}^2k^2)^3},
\end{equation}
%EQUAITON<<<
where $M_0$ and $\bar{\rho}$ denote constituent-quark mass at zero virtuality and average (anti)instanton size, respectively. From Ref.~\cite{Nam:2006ng}, $M_0$ is determined as about $350$ MeV for $1/\bar{\rho}\approx600$ MeV, to reproduce the pion weak-decay constant $F_\pi$.

\begin{figure}[t]
\begin{tabular}{cc}
\includegraphics[width=8.5cm]{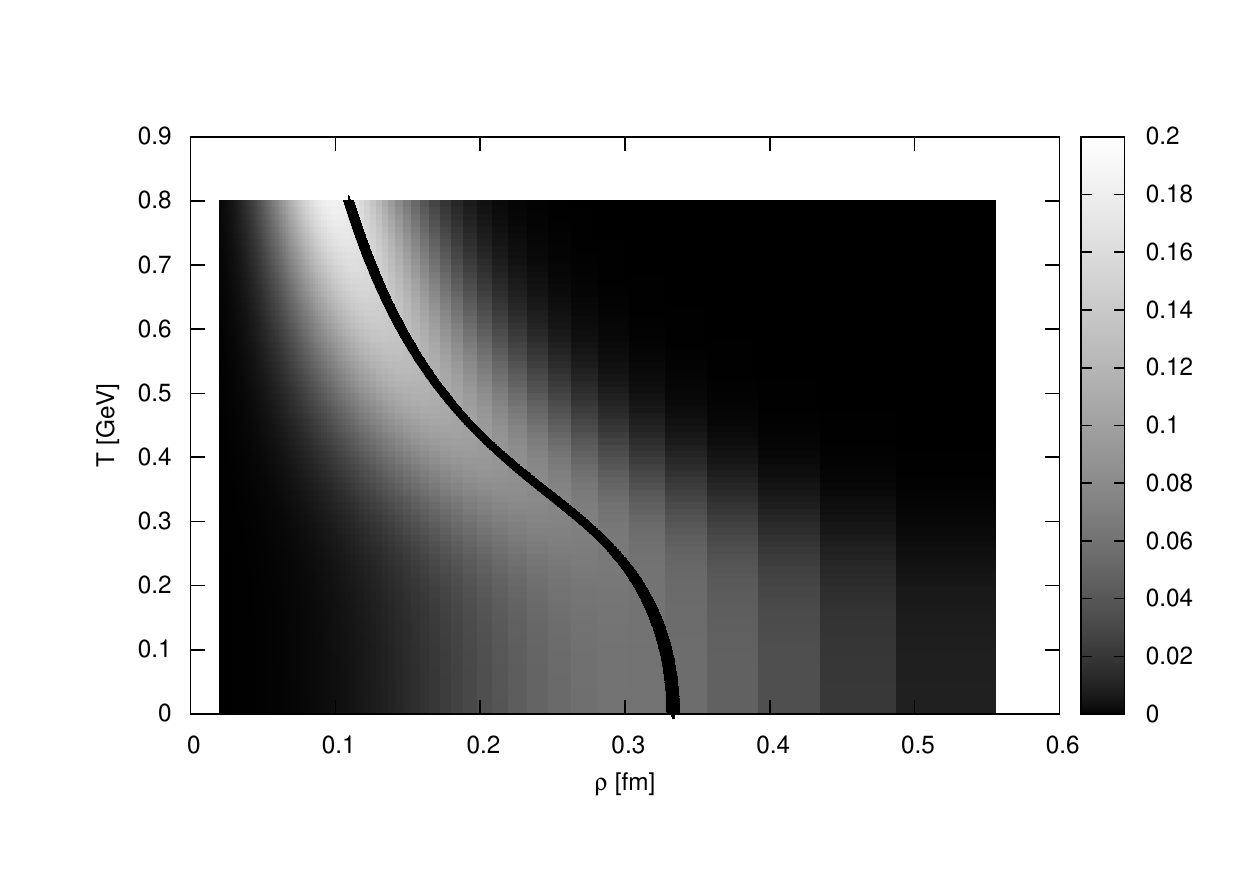}
\includegraphics[width=8.5cm]{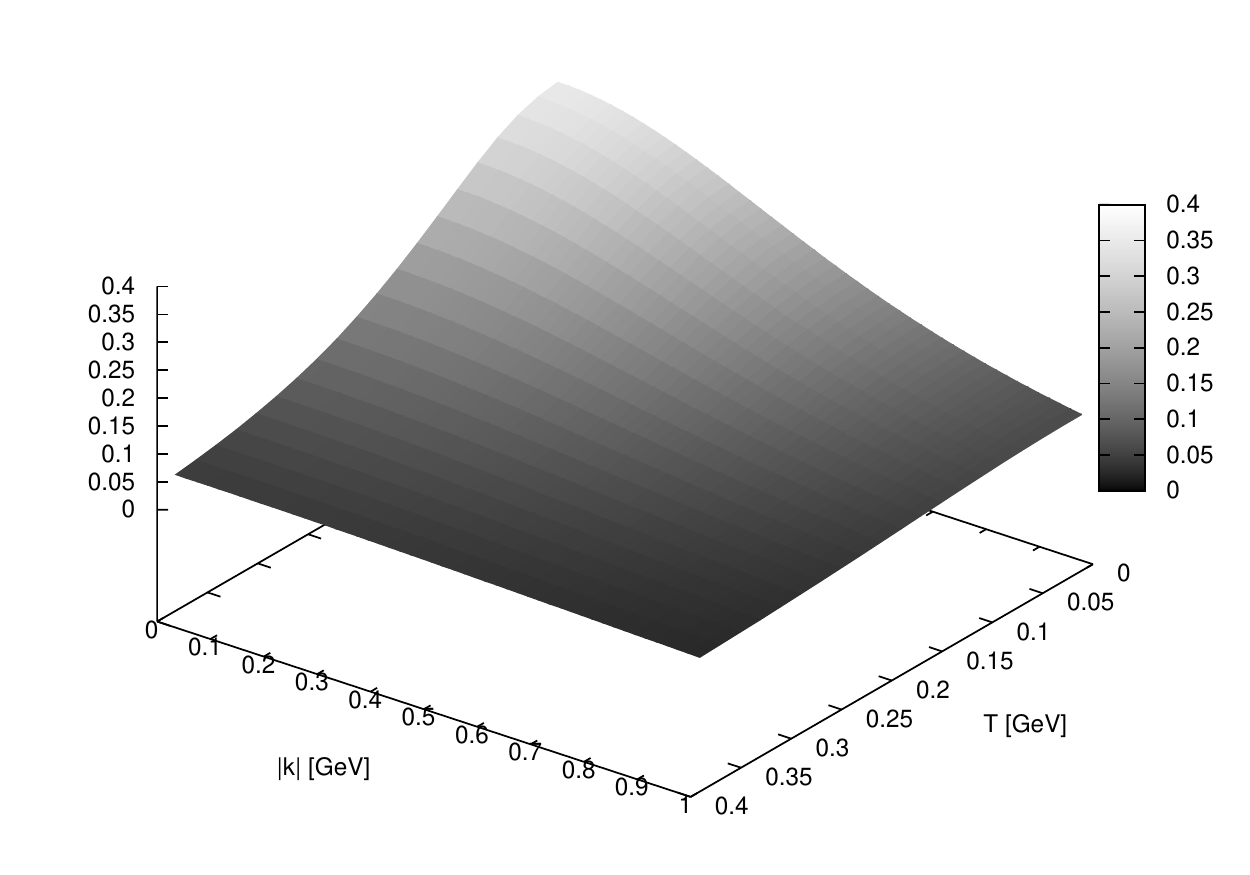}
\end{tabular}
\caption{Left: $d_N(\rho, T)$ on the $\rho$-$T$ plane. The trajectory on the curve represents $\bar{\rho}(T)$ in Eq.~(\ref{eq:rho}). Right: Momentum-temperature dependent effective quark mass in Eq.~(\ref{eq:momo}).}       
\label{FIG12}
\end{figure}

Now, we want to explain briefly how to modify $\bar{\rho}$ and $M_0$ as a function of $T$, using the caloron solution. Details can be found in Ref.~\cite{Nam:2009nn}.  An instanton distribution function for arbitrary $N_c$ and $N_f$ can be written with a Gaussian suppression factor as a function of $T$ and an arbitrary instanton size $\rho$ for pure-glue QCD~\cite{Diakonov:1988my}:
%EQUATION>>>
\begin{equation}
\label{eq:IND}
d(\rho,T)=\underbrace{C_{N_c}\,\Lambda^b_{\mathrm{RS}}\,
\hat{\beta}^{N_c}}_\mathcal{C}\,\rho^{b-5}
\exp\left[-(A_{N_c}T^2
+\bar{\beta}\gamma n\bar{\rho}^2)\rho^2 \right].
\end{equation}
%EQUATION<<<
We note that the CP-invariant vacuum was taken into account in Eq.~(\ref{eq:IND}), and we assumed the same analytical form of the distribution function for both the instanton and anti-instanton. Note that the instanton packing fraction $n\equiv1/\bar{R}^4$ and $\bar{\rho}$ have been taken into account as functions of $T$ implicitly. We also assigned the constant factor in the right-hand-side of the above equation as $\mathcal{C}$ for simplicity. The abbreviated notations are also given as:
%EQUATION>>>
\begin{eqnarray}
\label{eq:PARA}
\hat{\beta}&=&-b\ln[\Lambda_\mathrm{RS}\rho_\mathrm{cut}],\,\,\,\,
\bar{\beta}=-b\ln[\Lambda_\mathrm{RS}\langle R\rangle],\,\,\,
C_{N_c}=\frac{4.60\,e^{-1.68\alpha_{\mathrm{RS}} Nc}}{\pi^2(N_c-2)!(N_c-1)!},
\cr
A_{N_c}&=&\frac{1}{3}\left[\frac{11}{6}N_c-1\right]\pi^2,\,\,\,\,
\gamma=\frac{27}{4}\left[\frac{N_c}{N^2_c-1}\right]\pi^2,\,\,\,\,
b=\frac{11N_c-2N_f}{3}.
\end{eqnarray}
%EQUAITON<<<
Note that we defined the one-loop inverse charge $\hat{\beta}$ and $\bar{\beta}$ at certain phenomenological cutoff $\rho_\mathrm{cut}$ and $\langle R\rangle\approx\bar{R}$. $\Lambda_{\mathrm{RS}}$ denotes a scale, depending on a renormalization scheme, whereas $V_3$ for the three-dimensional volume. Using the instanton distribution function in Eq.~(\ref{eq:IND}), we can compute the average value of the instanton size $\bar{\rho}^2$ straightforwardly as follows~\cite{Schafer:1996wv}:
%EQUATION>>>
\begin{equation}
\label{eq:rho}
\bar{\rho}^2(T)
=\frac{\int d\rho\,\rho^2 d(\rho,T)}{\int d\rho\,d(\rho,T)}
=\frac{\left[A^2_{N_c}T^4
+4\nu\bar{\beta}\gamma n \right]^{\frac{1}{2}}
-A_{N_c}T^2}{2\bar{\beta}\gamma n},
\end{equation}
%EQUATION<<<
where $\nu=(b-4)/2$. It can be easily shown that Eq.~(\ref{eq:rho}) satisfies the  following asymptotic behaviors~\cite{Schafer:1996wv}:
%EQUATION>>>
\begin{equation}
\label{eq:asym}
\lim_{T\to0}\bar{\rho}^2(T)=\sqrt{\frac{\nu}{\bar{\beta}\gamma n}},
\,\,\,\,
\lim_{T\to\infty}\bar{\rho}^2(T)=\frac{\nu}{A_{N_c}T^2}.
\end{equation}
%EQUATION<<<
Here, the second relation of Eq.~(\ref{eq:asym}) indicates a correct scale-temperature behavior at high $T$, i.e., $1/\bar{\rho}\approx\Lambda\propto T$. Substituting Eq.~(\ref{eq:rho}) into Eq.~(\ref{eq:IND}), the caloron distribution function can be evaluated further:
%EQUATION>>>
\begin{equation}
\label{eq:dT}
d(\rho,T)=\mathcal{C}\,\rho^{b-5}
\exp\left[-\mathcal{F}(T)\rho^2 \right],\,\,\,\,
\mathcal{F}(T)=\frac{1}{2}A_{N_c}T^2+\left[\frac{1}{4}A^2_{N_c}T^4
+\nu\bar{\beta}\gamma n \right]^{\frac{1}{2}}.
\end{equation}
%EQUATION<<<
The instanton packing fraction $n$ can be computed self-consistently, using the following equation:
%EQUATION>>>
\begin{equation}
\label{eq:NOVV}
n^\frac{1}{\nu}\mathcal{F}(T)=\left[\mathcal{C}\,\Gamma(\nu) \right]^\frac{1}{\nu},
\end{equation}
%EQUATION<<<
where we replaced $NT/V_3\to n$, and $\Gamma(\nu)$ stands for the $\Gamma$-function with an argument $\nu$. Note that $\mathcal{C}$ and $\bar{\beta}$ can be determined easily using Eqs.~(\ref{eq:rho}) and (\ref{eq:NOVV}), incorporating the vacuum values for $n\approx(200\,\mathrm{MeV})^4$ and $\bar{\rho}\approx(600\,\mathrm{MeV})^{-1}$: $\mathcal{C}\approx9.81\times10^{-4}$ and $\bar{\beta}\approx9.19$. Finally, in order for estimating the $T$-dependence of $M_0$, one needs to consider the normalized distribution function, defined as follows:
%EQUATION>>>
\begin{equation}
\label{eq:NID}
d_N(\rho,T)=\frac{d(\rho,T)}{\int d\rho\,d(\rho,T)}
=\frac{\rho^{b-5}\mathcal{F}^\nu(T)
\exp\left[-\mathcal{F}(T)\rho^2 \right]}{\Gamma(\nu)}.
\end{equation}
%EQUATION<<<
Here, the subscript $N$ denotes the normalized distribution. For brevity, we want to employ the large-$N_c$ limit to simplify the expression for $d_N(\rho,T)$. In this limit, as understood from Eq.~(\ref{eq:NID}), $d_N(\rho,T)$ can be approximated as a $\delta$-function: 
%EQUATION>>>
\begin{equation}
\label{eq:NID2}
\lim_{N_c\to\infty}d_N(\rho,T)=\delta[{\rho-\bar{\rho}(T)}].
\end{equation}
%EQUATION<<<
The trajectory of this $\delta$-function projected on the $\rho$-$T$ plane is depicted in the left panel of Figure~\ref{FIG12} by the thick-solid line. Using Eq.~(\ref{eq:NID2}), we can write finally the ($|k|,T$)-dependent $M$ as follows:
%EQUATION>>>
\begin{equation}
\label{eq:momo}
M_k=M_0\left[\frac{\sqrt{n(T)}\,\bar{\rho}^2(T)}
{\sqrt{n(0)}\,\bar{\rho}^2(0)}\right]
\left[\frac{2}{2+k^2\,\bar{\rho}^2(T)} \right]^2,
\end{equation}
%EQUATION<<<
where we used $M_0\approx350$ MeV at $T\to0$. Note that $\bar{\rho}$ and $n$ are functions of $T$ in Eq.~(\ref{eq:momo}). We show the momentum-temperature dependent effective quark mass in Eq.~(\ref{eq:momo}) in the right panel of Figure~\ref{FIG12}. As shown there, the effective quark mass is a smoothly decreasing function of momentum as well as temperature as expected.
%FIGURE>>>
%FIGURE<<<

Considering all the ingredients discussed above and plugging Eq.~(\ref{eq:PRO}) into Eq.~(\ref{eq:EC1}), we obtain the following expression for the transverse components of $\sigma$ with $(\mu,\nu)=(1,1)$ or $(2,2)$ and $F_{12}=B_0$ in $z$ direction:
%EQUATION>>>
\begin{eqnarray}
\label{eq:EC2}
\sigma_{\mu\nu}(p)
&=&\sum_f\frac{4e^2_fN_c\delta_{\mu\nu}}{w_p}\int\frac{d^4k}{(2\pi)^4}\left[\frac{\frac{k(k+p)}{2}+M^2_k-2\tilde{M}^2_k(e_fB_0)^2}{[k^2+M^2_k][(k+p)^2+M^2_k]}\right].
\end{eqnarray}
%EQUAITON<<<
For the longitudinal one $(\mu,\nu)=(3,3)$, one removes the term proportional to $B^2_0$ in Eq.~(\ref{eq:EC2}). Since we are interested in the finite $T$ case, we introduce the fermionic Matsubara formula for Eq.~(\ref{eq:EC2}), then have the following expression:
%EQUATION>>>
\begin{eqnarray}
\label{eq:EC3}
\sigma_{\mu\nu}(p)&=&
\sum_f\frac{8e^2_fN_cT\delta_{\mu\nu}}{w_p}\sum_n\int\frac{d^3\bm{k}}{(2\pi)^3}\left[\frac{\frac{\bm{k}^2+w_k(w_k+w_p)}{2}+M^2_{\bm{k}}-2\tilde{M}^2_{\bm{k}}(e_fB_0)^2}{[\bm{k}^2+w^2_k+M^2_{\bm{k}}][\bm{k}^2+(w_k+w_p)^2+M^2_{\bm{k}}]}\right].
\end{eqnarray}
%EQUAITON<<<
Here, $n\ge0$ for  representing the retarded Green function. For definiteness, we made use of $p=(\vec{0},w_p)$ in deriving Eq.~(\ref{eq:EC3}). Note that we have simplified the mass functions in Eq.~(\ref{eq:EC3}) by replacing the four momentum into three one, $k\to\bm{k}$, in comparison to Eq.~(\ref{eq:EC2}):
%EQUATION>>>
\begin{equation}
\label{eq:MASS2}
M_{\bm{k}}\approx M_0(T)\,F_{\bm{k}}^2,\,\,
F_{\bm{k}}=\frac{2}{2+\bar{\rho}^2(T)\,\bm{k}^2},\,\,
\tilde{M}_{\bm{k}}\approx-\frac{8M_0(T)\,\bar{\rho}^2(T)}{[2+\bar{\rho}^2(T)\,\bm{k}^2]^3},
\end{equation}
%EQUAITON<<<
where $M_0(T)$ and $\bar{\rho}(T)$ are given in Eqs.~(\ref{eq:momo}) and (\ref{eq:NID2}). We verified that this simplification does not make considerable changes in the numerical results, and make the analytical and numerical calculations much easier. Note that the fermionic Matsubara frequency for the loop, $w_k$ is defined with the finite relaxation time $\tau$ for computing the finite electrical  conductivity as in Ref.~\cite{Kerbikov:2012vp}:
%EQUATION>>>
\begin{equation}
\label{eq:WK1}
w_k=w_n\left(1+\frac{1}{2\tau |w_n|} \right).
\end{equation}
%EQUAITON<<<
Here, $w_n=(2n+1)\pi T$ and $n\in\mathcal{I}$. We want to make an assumption: It is clear that the term $1/(2|w_n|)$ in Eq.~(\ref{eq:WK1}) becomes its maximum and most effective when $n=0$. Hence, it is reasonable to replace $|w_n|\to |w_0|=\pi T$ in the parentheses in the right-hand-side of Eq.~(\ref{eq:WK1}). Moreover, if we consider $w_p\approx 2\pi T$, $w_k$ in Eq.~(\ref{eq:WK1}) can be simplified further as follows:
%EQUATION>>>
\begin{equation}
\label{eq:WK2}
w_n\left(1+\frac{1}{2\pi T\tau} \right)\approx w_n\left(1+\frac{1}{w_p\tau} \right).
\end{equation}
%EQUAITON<<<
In what follows, we will use Eq.~(\ref{eq:WK2}) rather than Eq.~(\ref{eq:WK1}) for the Matsubara frequency in the calculations. 

The summations over $n$ in Eq.~(\ref{eq:EC3}) can be developed neatly as follows:
%EQUATION>>>
\begin{eqnarray}
\label{eq:SUM}
\sum_n\frac{T}{[\eta^2w^2_n+E^2_{\bm{k}}]}&=&\frac{1}{4\eta E_{\bm{k}}}
\mathrm{tanh}\left(\frac{E_{\bm{k}}}{2\eta T} \right),
\cr
\sum_n\frac{T}{[\eta^2w^2_n+E^2_{\bm{k}}]^2}&=&
\frac{1}{16\eta^2T E^3_{\bm{k}}}
\mathrm{sech}^2\left(\frac{E_{\bm{k}}}{2\eta T} \right)
\left[\eta T \mathrm{sinh}\left(\frac{E_{\bm{k}}}{\eta T} \right)-E_{\bm{k}}\right],
\end{eqnarray}
%EQUAITON<<<
where we have defined a notation as $\eta\equiv 1+1/(w_p\tau)$. Combining Eqs.~(\ref{eq:SIGMA}), (\ref{eq:WK2}), and (\ref{eq:SUM}), we have the following final expression for $\sigma$ for $w_p\to0$:
%EQUATION>>>
\begin{eqnarray}
\label{eq:EC4}
\sigma_\perp&=&\sum_fe^2_fN_c\tau
\Bigg\{\int\frac{d^3\bm{k}}{(2\pi)^3}
F^2_{\bm{k}}(\bm{k}^2)\left[
\frac{\mathrm{tanh}\left(\pi \tau E_{\bm{k}} \right)}{E_{\bm{k}}}\right] 
\cr
&+&\frac{\tau\pi}{2}\int\frac{d^3\bm{k}}{(2\pi)^3}
F^2_{\bm{k}}(\bm{k}^2)\left[\frac{\mathrm{sech}^2\left(\pi\tau E_{\bm{k}}\right)}{E^3_{\bm{k}}}
\left[\frac{\mathrm{sinh}\left(2\pi\tau E_{\bm{k}} \right)}{2\pi\tau}-E_{\bm{k}}\right]
\right]\left[M^2_{\bm{k}}-4\tilde{M}^2_{\bm{k}}(e_fB_0)^2 \right]\Bigg\}.
\end{eqnarray}
%EQUATION>>>
The energy of the quark is given by $E_{\bm{k}}=(\bm{k}^2+M^2_{\bm{k}})^{1/2}$. We note that we inserted $F^2_{\bm{k}}(\bm{k}^2)$ in the integrals in Eq.~(\ref{eq:EC4}) to tame the UV divergence smoothly in integrating over $\bm{k}$, instead of setting a three-dimensional cutoff. $\sigma_\parallel$ can be easily obtained by putting $B_0=0$ in Eq.~(\ref{eq:EC4}).

Finally in this Section, we want to discuss briefly about the chiral phase transition of the present model. One can explore the chiral  structure via the chiral (quark) condensate $\langle\bar{q}q\rangle$ within the present model~\cite{Nam:2009nn,Nam:2010mh}:
%EQUATION>>>
\begin{equation}
\label{eq:CC}
\langle\bar{q}q\rangle=-N_cN_f\int\frac{d^3\bm{k}}{(2\pi)^3}
\frac{M^2_{\bm{k}}}{E_{\bm{k}}}
\left[\frac{1-\exp\left(-\frac{3E_{\bm{k}}}{T} \right)}
{1+\exp\left(-\frac{3E_{\bm{k}}}{T} \right)} \right].
\end{equation}
%EQUAITON<<<
From the numerical results given in Ref.~\cite{Nam:2010mh}, the chiral condensate turns out to be a smoothly decreasing function of $T$ which is rather different from the universal class pattern of the chiral phase transition, i.e. there appears the second-order phase transition for the SU(2) light-flavor sector in the chiral limit $m_q\to0$. We note that, if the meson-loop correction (MLC), which relates to the $1/N_c$ correction, is taken into account, the correct second-oder phase transition was observed for the vanishing current-quark mass, giving $T_c\approx160$ MeV~\cite{Nam:2010mh}. Since we have not taken into account the MLC contribution in the present work, the chiral phase transition structure of the present theoretical framework is distinctive from that of the reality so that the inclusion of MLC can change the numerical results for $\sigma$ to a certain extent. We, however, note that the MLC contribution is ineffective for VCC $\sim\langle J_\mu J_\nu\rangle$ in the chiral limit in general~\cite{Goeke:2007bj}. Hence, it is rather safe to ignore the MLC contribution for the electrical conductivity as in the present work, despite of the problematic issue on the chiral structure at finite $T$ as discussed above. We would like to leave this interesting task for the inclusion of the MLC contribution for the future.

%--------------------------------------------------
\section{Numerical results and discussions}
\begin{figure}[t]
\begin{tabular}{cc}
\includegraphics[width=8.5cm]{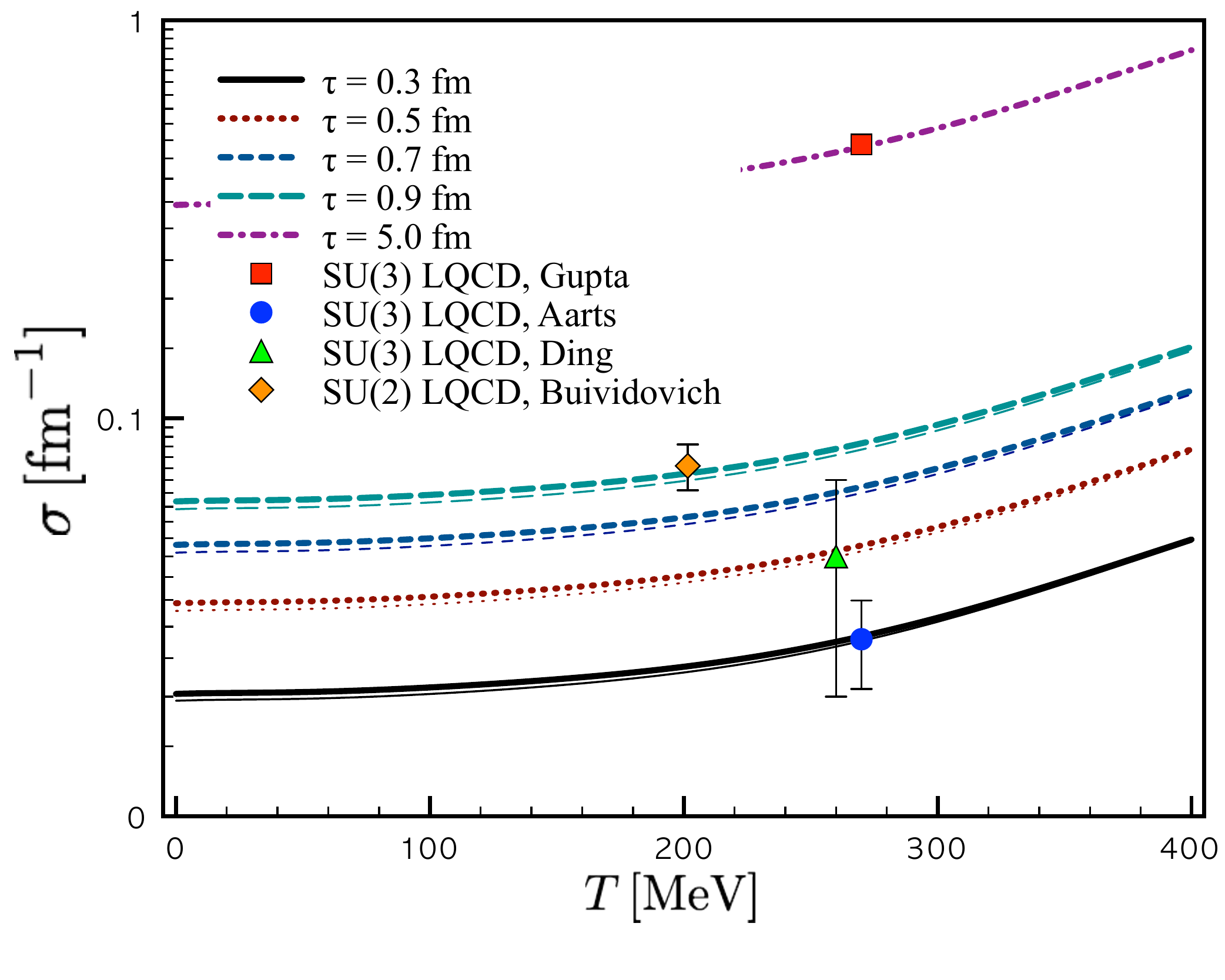}
\includegraphics[width=8.5cm]{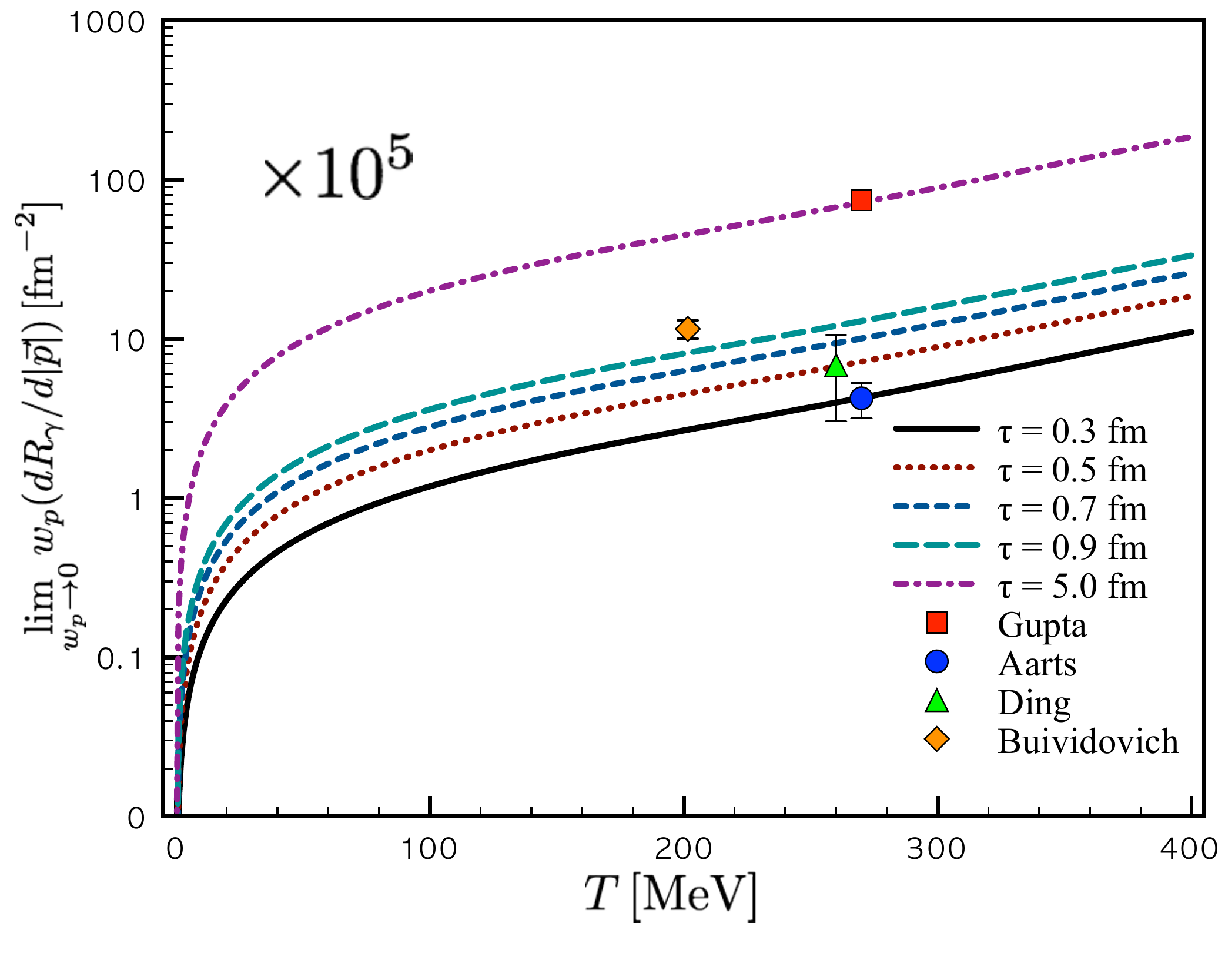}
\end{tabular}
\caption{(Color online) Left: electrical  conductivity $\sigma$ as functions of temperature $T$ for different relaxation times $\tau=(0.3\sim0.9)$ fm. The thick line indicate the case with $B_0=0$ ($\sigma_\parallel$), whereas the thin ones for $B_0=m^2_\pi\times10$ ($\sigma_\perp$), in Eq.~(\ref{eq:EC4}). SU($N_c$) LQCD estimations are taken from Refs.~\cite{Gupta:2003zh} (Gupta), ~\cite{Aarts:2007wj} (Aarts),~ \cite{Ding:2010ga} (Ding), and ~\cite{Buividovich:2010tn} (Buividovich). Right: Soft photon emission rate in Eq.~(\ref{eq:SPER}), $\mathcal{R}_\gamma$ in Eq.~(\ref{eq:SPER}) as functions of $T$ for different $\tau$'s in the same manner with the left panel.  For all the curves in the left and right panels, we have chosen $T_c\approx180$ MeV as a trial.}        
\label{FIG34}
\end{figure}

%--------------------------------------------------
In this Section, we present the numerical results for $\sigma$ and make comparisons with other theoretical results mainly from the LQCD simulation data. In the left panel of Figure~\ref{FIG34}, we show the numerical results of $\sigma_\perp$ (thick) and $\sigma_\parallel$ (thin) for different $\tau$ values, $\tau=(0.1,0.3,0.5,0.9)$ fm in (solid, dot, dash, dot-dash) lines, respectively, as functions of $T$. As for the external magnetic filed, we set it to be $B_0=m^2_\pi\times10$ as a trial. Note that this value of $B_0$ is still far stronger than that observed in the RHIC experiment~\cite{Voloshin:2008jx,Voloshin:2004vk,AA:2009txa,AA:2009uh}. From the figure, it turns out that $\sigma$ is a rapidly increasing function of $T$ and show the obvious increases beyond $T\approx200$ MeV. By comparing those cases with and without $B_0$, we also find that the effect from the external magnetic field is considerably small and only relatively effective in the low-$T$ region $T\lesssim200$ MeV. Hence, we can conclude from the present model that $\sigma_\perp\approx\sigma_\parallel\equiv\sigma$. For reference, we list $\sigma$ values for some typical temperatures in Table~\ref{TABLE1}, in which one can easily see that $\sigma$ is almost linear for $T\lesssim150$ MeV, and increases monotonically after that temperature. For instance, at $T_c\approx180$ MeV, which is close to the transition temperature of QCD, calculated from the full LQCD simulations~\cite{Maezawa:2007fd,Ali Khan:2000iz}, we have $\sigma=(0.023,0.039,0.054,0.070)\,\mathrm{fm}^{-1}$ for $\tau=(0.1,0.3,0.5,0.9)$ fm. It is worth mentioning that, in the recent LQCD simulations~\cite{Aoki:2006br,Aoki:2009sc,Borsanyi:2010bp,Bazavov:2011nk}, the transition temperature was obtained as $T_c\approx155$ MeV, which is lower than those from Refs.~\cite{Maezawa:2007fd,Ali Khan:2000iz}. At $T_c\approx155$ MeV for instance, we have $\sigma=(0.022,0.037,0.052,0.067)\,\mathrm{fm}^{-1}$ for $\tau=(0.1,0.3,0.5,0.9)$ fm. Hence, we conclude that there is only small change in $\sigma$ $(\sim\mathrm{a\,few\,}\%)$ for different $T_c$ values below $200$ MeV as shown in the left panel of Figure~\ref{FIG34}.

%TABLE>>>
\begin{table}[t]
\begin{tabular}{c|c|c|c|c|c}
&\hspace{0.5cm}$T=0$\hspace{0.5cm}
&$T=100$ MeV&$T=200$ MeV&$T=300$ MeV&$T=400$ MeV\\
\hline
$\tau=0.3$ fm&$0.020$&$0.021$&$0.024$&$0.031$&$0.049$\\
$\tau=0.5$ fm&$0.034$&$0.036$&$0.040$&$0.053$&$0.083$\\
$\tau=0.7$ fm&$0.048$&$0.050$&$0.056$&$0.074$&$0.116$\\
$\tau=0.9$ fm&$0.062$&$0.064$&$0.072$&$0.095$&$0.149$\\
\end{tabular}
\caption{Typical values of $\sigma$ in Eq.~(\ref{eq:EC4}) [fm$^{-1}$] at $B_0=0$ 
for different $T$ and $\tau$ values.}
\label{TABLE1}
\end{table}
%TABLE>>>

For practical applications as in the LQCD simulations~\cite{Gupta:2003zh,Aarts:2007wj}, it is quite convenient to parameterize the present numerical results for $\sigma$ as follows:
%EQUATION>>>
\begin{equation}
\label{eq:PARAEC}
\sigma(T)=C_\mathrm{EM}\sum_{m=1}\mathcal{C}_m T^m,\,\,\,\,\frac{\mathcal{C}_m}{\mathrm{fm^{m-1}}}\in\mathcal{R}.
\end{equation}
%EQUAITON<<<
Here, we define $C_\mathrm{EM}=\sum_fe^2_f\approx0.051$ for the SU(2) light-flavor sector. Obtained coefficients up to $m=3$ are given in Table~\ref{TABLE2}. As shown in the coefficients, the electrical  conductivity can be parameterized almost linearly with respect to $T$, i.e. $|\mathcal{C}_{2,3}|\sim0$. Hence, we can approximate them as $\sigma/T\approx (0.46,0.77,1.08,1.39)\,C_\mathrm{EM}$ for $\tau=(0.3,0.5,0.7,0.9)$ fm, respectively.
%TABLE>>>
\begin{table}[b]
\begin{tabular}{c||c|c|c|c}
&$\tau=0.3$ fm&$\tau=0.5$ fm&$\tau=0.7$ fm&$\tau=0.9$ fm\\
\hline
$\mathcal{C}_1$&$0.46$&$0.77$&$1.08$&$1.39$\\
$\mathcal{C}_2$ [fm] &$4.00\times10^{-6}$&$6.66
\times10^{-6}$&$9.33\times10^{-6}$&$1.20\times10^{-6}$\\
$\mathcal{C}_3$ [fm$^2$]&$-4.87\times10^{-5}$&$-4.87
\times10^{-6}$&$-4.88\times10^{-5}$&$-4.88\times10^{-5}$\\
\end{tabular}
\caption{The coefficients $\mathcal{C}_{1,2,3}$ for Eq.~(\ref{eq:PARAEC}) for different $\tau$ values.}
\label{TABLE2}
\end{table}
%TABLE>>>

In Refs.~\cite{Gupta:2003zh} and~\cite{Aarts:2007wj}, employing the SU(3) quenched LQCD simulations, it was estimated that $\sigma/T=7C_\mathrm{EM}$ for $1.5<T/T_c<3$ and $\sigma/T=(0.4\pm0.1)C_\mathrm{EM}$ for $T/T_c\approx1.5$, respectively. Note that there is one order difference between these $\sigma$ values, although the temperature ranges are not overlapped. In the left panel of Figure~\ref{FIG34}, we depict these two LQCD values from Ref.~\cite{Gupta:2003zh} (square) and Ref.~\cite{Aarts:2007wj} (circle), using $T_c\approx180$ MeV as a trial, although the transition temperatures are slightly higher than this value in general in the quenched LQCD simulations. It is shown that the data point from Ref.~\cite{Aarts:2007wj} is well consistent with our numerical result for $\tau\approx0.3$ fm. On the contrary, the data point from Ref.~\cite{Gupta:2003zh} for $T=270$ MeV is much larger than ours for $\tau=(0.3\sim0.9)$ fm. We verified that, in order to reproduce it, $\tau$ becomes about $5$ fm in our model calculation as shown in the left panel of Figure~\ref{FIG34} in the dot-dash line. In Ref.~\cite{Tuchin:2010vs}, the characteristic $\tau$ was estimated using Ref.~\cite{Gupta:2003zh}, resulting in about $(2.2\,T/T_c)$ fm with a conservative estimate of the QGP medium size. Taking $T\approx270\,\mathrm{MeV}=T_c\times1.5$ MeV, it is given that $\tau\approx5$ fm, which is in good consistency with our model estimation as shown in the left panel of Figure~\ref{FIG34}. Similarly, at $T\approx1.45\,T_c$, it was suggested that $\sigma/T\approx (1/3\sim1)C_\mathrm{EM}$ in Ref.~\cite{Ding:2010ga}. If we choose $T_c\approx180$ MeV again, this result provides that $\sigma=(0.022\sim0.067)\,\mathrm{fm}^{-1}$, which is drawn in the left panel of Figure~\ref{FIG34} (triangle) and it corresponds to $\tau\approx(0.3\sim0.7)$ fm in comparison with our results. Moreover, the typical time scale of $\tau$ was given by $\tau T=0.5$, giving $\tau\approx0.38$ fm at $T\approx1.45\,T_c$~\cite{Ding:2010ga}. Interestingly, this time scale is quite compatible with ours. In those LQCD data, it should be noticed that the unrenormalized vector currents for the Kubo formula $\sim\langle J_\mu J_\nu\rangle$ were employed in Refs.~\cite{Gupta:2003zh,Aarts:2007wj}, whereas Ref.~\cite{Ding:2010ga} made use of the renormalized one. Hence, one may expect systematic errors for $\sigma$ given in Refs.~\cite{Gupta:2003zh,Aarts:2007wj}. To estimate those errors are, however, clearly beyond our scope.

In Ref.~\cite{Buividovich:2010tn}, the quenched SU(2) LQCD was performed, and the electrical  conductivity was also estimated as $\sigma=(0.076\pm0.010)\,\mathrm{fm}^{-1}$ at $T=350$ MeV with the transition temperature $\sim313$ MeV, taking into account $T=1.12\,T_c$. To be consistent with others using $T_c=180$ MeV as above, we depict the data point of Ref.~\cite{Buividovich:2010tn} at $T=1.12\times180\,\mathrm{MeV}\approx202$ MeV in the left panel of Figure~\ref{FIG34} (diamond), although it was evaluated at $T=350$ MeV. Being different from other LQCD data, Ref.~\cite{Buividovich:2010tn} presented the longitudinal and transverse components of $\sigma$ separately in the presence of the external magnetic field. Those LQCD data showed that $\sigma_\perp\approx\sigma_\parallel$ beyond $T_c$ for arbitrary values of $B_0$, whereas $\sigma_\perp\ne\sigma_\parallel$ at $T=0$ and the difference between them is enhanced by increasing $B_0$. Qualitatively, this observation of the LQCD results are consistent with ours as indicated by the thick and thin lines in the left panel of Figure~\ref{FIG34}. In our calculations, $\sigma_\parallel$ (thin) is smaller than $\sigma_\perp$, mainly due to that the negative sign in front of the term $\propto(e_fB_0)^2$ in Eq.~(\ref{eq:EC4}) in the vicinity of $T\approx0$. On the contrary, $\sigma_\parallel$ becomes larger than $\sigma_\perp$ at $T=0$ in Ref.~\cite{Buividovich:2010tn}. 

However, we note that, if we consider the finite number-difference between the instanton and anti-instanton numbers, i.e.  $\Delta N\equiv N_I-N_{\bar{I}}\ne0$~\cite{Nam:2009jb}, which are not not taken into account in the present work and corresponds to U$_\mathrm{A}$(1) QCD anomaly via the finite chiral quark density related to CME~\cite{Kharzeev:2004ey,Kharzeev:2007jp,Buividovich:2009wi,Fukushima:2008xe}, the conductivity increases with respect to the eternal magnetic field. In other words, in order to observe sizable values for the electrical  conductivity enhanced by the external magnetic fields, it is necessary to have U$_\mathrm{A}$(1) QCD anomaly explicitly in the model, in addition to the anomaly-independent terms~\cite{Nam:2009jb}. 

Beside the LQCD data, there were several theoretical estimations for the electrical  conductivity via effective approaches using the Green-function method~\cite{Kerbikov:2012vp} and ChPT~\cite{FernandezFraile:2009mi}. In Ref.~\cite{Kerbikov:2012vp}, the electrical  conductivity was computed for finite temperature and quark density, $T=100$ MeV and $\mu=400$ MeV, which may correspond to the future heavy-ion collision facilities, such as Facility for Antiprotons and Ions Research (FAIR) of GSI and Nuclotron-based Ion Collider fAcility (NICA) of JINR as noted there. By choosing $\tau=0.9$ fm, it was given that $\sigma\approx0.04\,\mathrm{fm}^{-1}$. Note that this value corresponds to $\tau=(0.5\sim0.7)$ in our results for zero density. In other words, by increasing the quark density, the electrical  conductivity decreases at a certain temperature, as expected.

%FIGURE>>>
%FIGURE<<<

Finally, we would like to estimate the (differential) soft photon ($w_p\to0$) emission rate from QGP for the dilepton decay rates which is related to the electrical  conductivity as follows~\cite{Gupta:2003zh,Ding:2010ga}: 
%EQUATION>>>
\begin{equation}
\label{eq:SPER}
\mathcal{R}_\gamma\equiv
\lim_{w_p\to0}w_p\frac{dR_\gamma}{d^3|\bm{p}|}=\frac{3\alpha_\mathrm{EM}}{2\pi^2}\sigma T.
\end{equation}
%EQUAITON<<<
The numerical results for $\mathcal{R}_\gamma$ in Eq.~(\ref{eq:SPER}) are given in the right panel of Figure~\ref{FIG34} for different $\tau$ values as in the left panel of Figure~\ref{FIG34}. We also depict the other theoretical estimations there, using Eq.~(\ref{eq:SPER}). It turns out that the value of $\mathcal{R}_\gamma$ increases rapidly in the vicinity of $T=0$. Beyond that, the slope of $\mathcal{R}_\gamma$ becomes relatively stable with respect to $T$. Although one can easily have numerical values for $\mathcal{R}_\gamma$ from the numbers given in Table~\ref{TABLE1} by multiplying $3\alpha_\mathrm{EM} T/(2\pi^2)$, for convenience for readers, $\mathcal{R}_\gamma$ for some temperatures are listed in Table~\ref{TABLE3}. 

%TABLE>>>
\begin{table}[h]
\begin{tabular}{c|c|c|c|c}
&$T=100$ MeV&$T=200$ MeV&$T=300$ MeV&$T=400$ MeV\\
\hline
$\tau=0.3$ fm&$1.181\times10^{-5}$&$2.700\times10^{-5}$
&$5.229\times10^{-5}$&$11.021\times10^{-5}$\\
$\tau=0.5$ fm&$2.024\times10^{-5}$&$4.500\times10^{-5}$
&$8.940\times10^{-5}$&$18.667\times10^{-5}$\\
$\tau=0.7$ fm&$2.811\times10^{-5}$&$6.297\times10^{-5}$
&$12.482\times10^{-5}$&$26.089\times10^{-5}$\\
$\tau=0.9$ fm&$3.600\times10^{-5}$&$8.100\times10^{-5}$
&$16.025\times10^{-5}$&$33.511\times10^{-5}$\\
\end{tabular}
\caption{Typical values of $\mathcal{R}_\gamma$ in Eq.~(\ref{eq:SPER}) [fm$^{-2}$] at $B_0=0$ for different $T$ and $\tau$ values.}
\label{TABLE3}
\end{table}
%TABLE>>>

%--------------------------------------------------
\section{Summary and conclusion}
%--------------------------------------------------
We have investigated the electrical  conductivity of quark matter at finite temperature and zero quark density $(T\ne0,\mu=0)$, using the dilute instanton-liquid model. The instanton parameters, i.e. the average instanton size and inter-(instanton) distance, were modified as functions of $T$ according to the caloron distribution. Employing the Kubo formula, we computed it in the presence of the external magnetic field in the order of $B_0\approx m^2_\pi\times10$. We compared our results with the LQCD data as well as effective theories. Important observations of the present work are listed as follows:
%ITEMIZE>>> 
\begin{itemize}
%---------------
\item The electrical conductivity is an increasing function of $T$ and the values for $\sigma$ depends on the relaxation time $\tau$ of the quark matter. Note that we modify the effective quark mass into a decreasing function of $T$ as well as $|\bm{k}|$, playing the role of a natural UV regulator and signaling the reduction of the nonperturbative QCD effects at finite $T$. Typically, we have $\sigma\approx(0.02\sim0.15)\,\mathrm{fm}^{-1}$ for $T=(0\sim400)$ MeV with the relaxation time $\tau=(0.3\sim0.9)$ fm. 
%---------------
\item
Recent theoretical estimations including the quenched LQCD data are well reproduced for $\tau=(0.3\sim0.7)\,\mathrm{fm}$ for a wide $T$ range. From the numerical calculations, it turns out that the one-order larger $\sigma$, estimated in the early LQCD data~\cite{Gupta:2003zh}, corresponds to $\tau\approx5$ fm, which is much larger than other values. This value $\tau\approx5$ fm is well consistent with a conservative estimate of the QGP-medium size.
%---------------
\item The external magnetic field provides only negligible effects on the electrical conductivity even for the very strong one in the order of $B_0\approx m^2_\pi\times10$. However, if we consider the U$_\mathrm{A}$(1) anomaly, which is the main source for CME and produced from the number difference between the instanton and anti-instanton in the present theoretical framework, this situation can be altered to a certain extent.
%---------------
\item Using the numerical results obtained, the electrical conductivity is parameterized for $T=(0\sim400)$ MeV with $\sigma=C_\mathrm{EM}(\mathcal{C}_1T+\mathcal{C}_2T^2+\mathcal{C}_3T^3+\cdots)$ and, in this parameterization, the coefficients for $T^2$ and $T^3$ are almost negligible in comparison to $\mathcal{C}_1$. As a result, we have $\sigma/T\approx (0.46,0.77,1.08,1.39)\,C_\mathrm{EM}$ for $\tau=(0.3,0.5,0.7,0.9)$ fm, respectively. These results are again well compatible with other theoretical estimations. 
%---------------
\end{itemize}
%ITEMIZE>>>

In order to understand QCD at extreme conditions, the transport coefficients for quark matter are very important physical quantities to be studied. In the present work, we have shown that the instanton model reproduced compatible results for the electrical conductivity with other theoretical calculations including the LQCD data, and estimated it as a function of $T$ for different relaxation times of the matter for the first time. Hence, it is quite promising and challenging to investigate other coefficients as well within in the same theoretical framework. Other transport coefficients such as the shear and bulk viscosities are under investigation, and related works will appear elsewhere.    
%--------------------------------------------------

%--------------------------------------------------
\end{document}